\documentclass[twocolumn,prl,amsmath,amssymb,showpacs]{revtex4-1}
\usepackage{epsf}      
\usepackage{graphicx}
\usepackage{color}
\usepackage{gensymb}
\usepackage{amsmath}
\usepackage{array}

\begin{document}

\title{Structural phase transition and its consequences on optical behavior of LaV$_{1-x}$Nb$_x$O$_4$}

\author{Hemanshu Dua}
\email{These authors have contributed equally to this work}
\affiliation{Department of Physics, Indian Institute of Technology Delhi, Hauz Khas, New Delhi-110016, India}
\author{Rishabh Shukla}
\email{These authors have contributed equally to this work}
\affiliation{Department of Physics, Indian Institute of Technology Delhi, Hauz Khas, New Delhi-110016, India}
\author{R. S. Dhaka}
\email{rsdhaka@physics.iitd.ac.in}
\affiliation{Department of Physics, Indian Institute of Technology Delhi, Hauz Khas, New Delhi-110016, India}

\date{\today}                          

\begin{abstract}
We present the structural, electronic, vibrational, and photoluminescence properties of polycrystalline LaV$_{1-x}$Nb$_x$O$_4$ ($x =$ 0--0.2) samples at room temperature. The substitution of Nb at the V site shows the fascinating structural and optical behavior due to their isoelectronic character and larger ionic radii of Nb$^{5+}$ as compared to the V$^{5+}$. The Rietveld refinement of x-ray diffraction patterns demonstrate that the $x=$ 0 sample exist in a monoclinic (P2$_1$/n) phase, whereas for the $x >$ 0, both monoclinic and scheelite-tetragonal (I4$_1$/a) phases co-exist in a certain proportion. Interestingly, a monotonous enhancement in the Raman spectral intensity with Nb substitution is correlated with the substitution induced increase in the scheelite-tetragonal phase. The x-ray absorption measurements reveal that the La ions exist in a trivalent oxidation state, while V and Nb cations possess 5+ oxidation state in tetrahedral coordination. Moreover, the Fourier-transform infrared (FTIR) spectra indicate that the Nb substitution give origin to some additional IR modes owing to the deformation of the VO$_4$$^{3-}$ tetrahedra and mixing of monoclinic and tetragonal phases. The photoluminescence measurements on these samples exhibit broadband spectra and their deconvolution designate the availability of more than one electron-hole pairs recombination center.
\end{abstract}

\maketitle

\section{\noindent ~Introduction}

Orthovanadates are an eminent class of compounds with technological and fundamental significance, which have been extensively utilized as luminescent materials, polarizers, catalysts, biological sensors, battery electrodes, alternatives in green technologies, etc. \cite{ZhaoCEC12, WangACR11, BulbulMCP16, YiJALCOM17, ZhangJAC13}. These compounds exhibit phenomenon of temperature and pressure-dependent structural phase transformation \cite{ErrandoneaPRB09, ErrandoneaPMS18}, which was debated for a long time among the researchers for its order type \cite{SmirnovPRB08}. In recent years, rare earth orthovanadates having a general formula of RVO$_4$ (R = rare-earth ion) have gained tremendous popularity for their useful applications in the area of solar cells \cite{KimNE15, ZahedifarJL13}, thin film phosphors \cite{YuAPA05}, photocatalysis \cite{HeJMCA11, MahapatraIECR07}, etc. For enhancement in the optical performance of the orthovanadate compounds, researchers predominantly follow three paths, which can be defined as the substitution of metal cations \cite{DolgosJSSC09}, coupling with the other metal oxides \cite{LinIJHE06}, and synthesis of these compounds with various novel routes \cite{FanJPCB06, JiaJPCB05, SelvenJCS09}. Primarily, the RVO$_4$ materials are known to crystallize into two polymorphs, namely monoclinic (m-) monazite-type (space group: P2$_1$/n, $Z$= 4) and tetragonal (t-) zircon-type (space group: I4$_1$/amd, $Z$= 4). Note that the space group P2$_1$/n is defined as a non-conventional setting of the standard P2$_1$/c space group (\#14) and can be obtained by changing the basis vector in the matrix transformation of P2$_1$/c \cite{RiceACB76}. In a monazite structure, R$^{3+}$ cations form an edge-sharing nonahedra (RO$_9$) with distorted VO$_4$ tetrahedra along the c-axis having four dissimilar V--O bonds \cite{RiceACB76}, while tetragonal structure possesses edge-sharing dodecahedra (RO$_8$) and undistorted VO$_{4}$ tetrahedral chains parallel to the c-axis with four identical V--O bonds \cite{ChakoumkosJSSC94}. It has been observed that the larger size of lanthanide ions (Ln$^{3+}$) favor monoclinic structure over tetragonal one due to a higher oxygen coordination number \cite{ChakoumkosJSSC94, MahapatraIECR07}. Hence, the La$^{3+}$ cations having the largest ionic radii in the Ln$^{3+}$ series generally crystallize into the thermodynamically stable monoclinic structure and a metastable tetragonal structure \cite{ShannonACB69, FanJPCB06}, while other lanthanide orthovanadates predominantly crystallize into the zircon-type tetragonal structure \cite{RoppJINC73}.

Interestingly, the LaVO$_4$ can also be crystallized into a stable tetragonal structure in the form of nanocrystals/nanowires via the solution method but stabilizing the tetragonal structure is always been a challenging task \cite{JiaJPCB05, FanJSSC07, RoppJINC73}. A combination of the experiments like temperature-dependent photoconductivity, emission measurements, and different empirical models were employed to explain the complete energy level diagram of lanthanide-doped LaVO$_4$ compounds \cite{KrumpelMSEB08}, which can be utilized to explain their luminescence properties and would help in the preparation of new luminescent materials. This is well known that the luminescence properties of lanthanide (Ln) doped rare-earth orthovanadates depend on the location of the 4$f$ energy levels of the Ln dopants with respect to the valence and conduction band of the parent compound \cite{KrumpelJPCM09}. The t-LaVO$_4$ exhibits prominent photoluminescence properties as compared to its m-counterpart because the t-LaVO$_4$ has four sigma bonds of angle 153$\degree$, which leads to the efficient transfer of energy whereas, in m-LaVO$_4$ the bond angles are smaller than t-LaVO$_4$, which results in the less effective transfer of energy \cite{ParkPB10}. The first-principles calculations for LaVO$_4$ polymorph performed using plane-wave pseudopotential method suggest that the m-LaVO$_4$ has an indirect bandgap of 3.5~eV, while the t-LaVO$_4$ has a direct bandgap of 3~eV \cite{SunJAP10}. Thus, the phase transformation from m- to t-structure leads to a significant increase in the photoluminescent properties of LaVO$_4$, which are attributed to the structural difference of m- and t-phases. 

In recent years, intensive research is being carried out in order to ameliorate the optical properties of LaVO$_4$ via substituting M$^+$(Li), M$^{2+}$ (Mg, Sr, etc.), M$^{3+}$ (Eu, Sm, etc.), and M$^{4+}$ (Sn, Zr, etc.) ions at La site \cite{JiaAPL04, RambabuMRB00, LiuMCP09, YuAPA05, YinML17, TyminskiJALCOM19, LiuAM07, OkramIC14}, however, enhancement in the performance with the substitution at V site still has room to explore for the researchers, where only a report of Mn$^{4+}$ cationic substitution induced catalytic properties are present in the literature \cite{VermaACAG01}. Therefore, in order to enhance the optical performance of the LaVO$_4$ compound with B-site substitution, we have chosen the isoelectronic Nb$^{5+}$ cations having a larger ionic radius of 0.48~\AA~as compared to V$^{5+}$ (0.355~\AA) ions \cite{ShannonACB69}. Moreover, the end member LaNbO$_4$ is well known for its multifunctional applications in phosphors and scintillators \cite{DingSSC18, YanOM06}, great tunability \cite{HsiaoJL07}, an efficient blue luminescence material when excited with UV/X-ray source \cite{BlaseCPL90}. Also, the LaNbO$_4$ manifests a second-order temperature-dependent structural phase transformation in a temperature range of 500$\pm$20$\rm^o$C from a fergusonite-monoclinic (f-m) phase (space group: I2/a, \#15) into a scheelite-tetragonal (s-t) phase (space group: I4$_1$/a, \#88) \cite{JianJACS97, HuseJSSC12}. Both the phases (f-m and s-t) of LaNbO$_4$ compound consist of the NbO$_4$ tetrahedra and LaO$_8$ dodecahedra, which reflects the twelve triangular lattices \cite{HuseJSSC12}. The s-t structure has all 4$\times$Nb-O equal bond-lengths including two sets of 4$\times$La-O bonds, whereas in the f-m phase two sets of Nb-O bonds appear with four pairs of La-O bond-lengths \cite{HuseJSSC12}. This scheelite tetragonal (s-t) phase is known as a close-packed (denser) polymorph with a volume reduction of nearly 10\% with respect to the standard zircon-type tetragonal phase \cite{ErrandoneaPRB09}. Moreover, a fergusonite to scheelite transformation is obtained from the cyclic rotation of axes with a transformation matrix and the long-axis (unique b-axis) in the fergusonite phase become the c-axis of the scheelite phase with a gradual change of $\beta$ from 94$\rm^o$ to 90$\rm^o$ \cite{ErrandoneaJPCM19, HuseJSSC12}. These tetragonal phases in the orthovanadates were known to possess the remarkable optical properties and hence being explored for their practical applications \cite{PanchalJAP11}. 

Therefore, in this paper we investigate the structural, electronic, vibrational, and photoluminescence properties of LaV$ _{1-x}$Nb$_x$O$_4$ ($x=$ 0--0.2). The Rietveld refinement of x-ray diffraction patterns demonstrates the monoclinic (P2$_1$/n) phase ($x=$ 0), and appearance of an additional scheelite-tetragonal (I4$_1$/a) phase for the $x>$ 0 samples. The room temperature x-ray absorption spectroscopy (XAS) measurements affirm the presence of trivalent La cations and a pentavalent oxidation state of V and Nb ions in tetrahedral coordination. The larger ionic radii of Nb$^{5+}$ as compared to V$^{5+}$ results in the structural phase transformation from a monoclinic to a scheelite-tetragonal. Moreover, an enhancement in the Raman spectral intensity is related to the increase in the tetragonal-phase with the Nb substitution at V site. Furthermore, additional infrared modes in the Fourier-transform infrared (FTIR) spectra were observed due to the Nb induced structural transformation. In the room temperature PL measurements, we found that the strongly overlapped spectra can be deconvoluted into six peaks, where each emission process involves the separate energy levels.

\section{\noindent ~Experimental}
The LaV$_{1-x}$Nb$_x$O$_4$ ($x=$ 0--0.2) samples were prepared via solid-state reaction method by mixing V$_2$O$_5$ (99.6$\%$, Sigma), Nb$_2$O$_5$ (99.99$\%$, Sigma) and La$_2$O$_3$ (99.99$\%$, Sigma) in the stoichiometric amount as initials, where La$_2$O$_3$ was pre-dried at 900$\degree$C for 6 hrs to remove moisture. The mixture was homogeneously ground for 8 hrs, followed by calcination at 1000$\rm^o$C for 17~hrs, then the obtained mixture was reground and sintered at 1200$\rm^o$C for 13~hrs. To improve the crystallinity of prepared samples, final sintering was done at 1250$\rm^o$C for 13~hrs. We have also sintered the Nb substituted samples ($x =$ 0.1 and 0.2) at 1450$\rm^o$C for 13~hrs to explore the sintering temperature-induced phase transformation. The powder x-ray diffraction (XRD) data of prepared polycrystalline LaNb$_x$V$ _{1-x}$O$_{4} $ ($x=$ 0--0.2) samples were recorded using CuK$_\alpha$ radiation ($\lambda$ = 1.5406~\AA) from Rigaku Ultima IV, Ri x-ray diffractometer. The Rietveld refinement of recorded XRD patterns were accomplished using FullProf software \cite{CarvajalPB93} with background fitted using linear interpolation between the data points. The x-ray absorption measurements were recorded at room temperature in the transmission mode at La-L$_2$, Nb-K and La-L$_3$+V-K edges using the scanning EXAFS beamline (BL-09), Indus-2, RRCAT, Indore, India. The energy-dispersive x-ray analysis (EDX) was performed with the Zeiss EVO 18 at an operating voltage of 20~keV inside a vacuum chamber. The Raman spectra were recorded with the Renishaw inVia confocal Raman microscope using a 532~nm laser with grating 2400~lines/mm, 20X objective, and a power of 0.05~mW. The FTIR spectra were recorded using the Thermo Scientific Nicolet iS50 FT-IR  spectrophotometer in the attenuated total reflection (ATR) mode within the range of 400--4000~cm$^{-1}$. This instrument has a separate assembly for the measurement of FTIR spectra in the ATR mode. In the ATR mode, a diamond crystal was used as an ATR crystal and the sample was in the fine powder form to ensure an intimate contact with the diamond crystal. First, we record the background spectrum for the ATR crystal and then measure the spectrum for each sample. In the manuscript, the presented FTIR spectra for each sample are corrected with this background spectrum of the diamond. The photoluminescence (PL) spectra were recorded using the Horiba Jobin Yvon; LabRAM HR evolution system equipped with a UV laser of the wavelength of 325~nm (He-Cd laser, 30~mW). The PL spectra were recorded with the 325~nm excitation laser source, a 40X objective, 2400 lines/mm grating, and 3~mW laser power. A charged coupled device (CCD) detector was used for the data collection. Finally, the diffuse reflectance spectroscopic measurements were conducted using a Shimadzu UV-2450 spectrophotometer at a slit width of 5~mm.

\begin{figure*}
\includegraphics[width=7.0in]{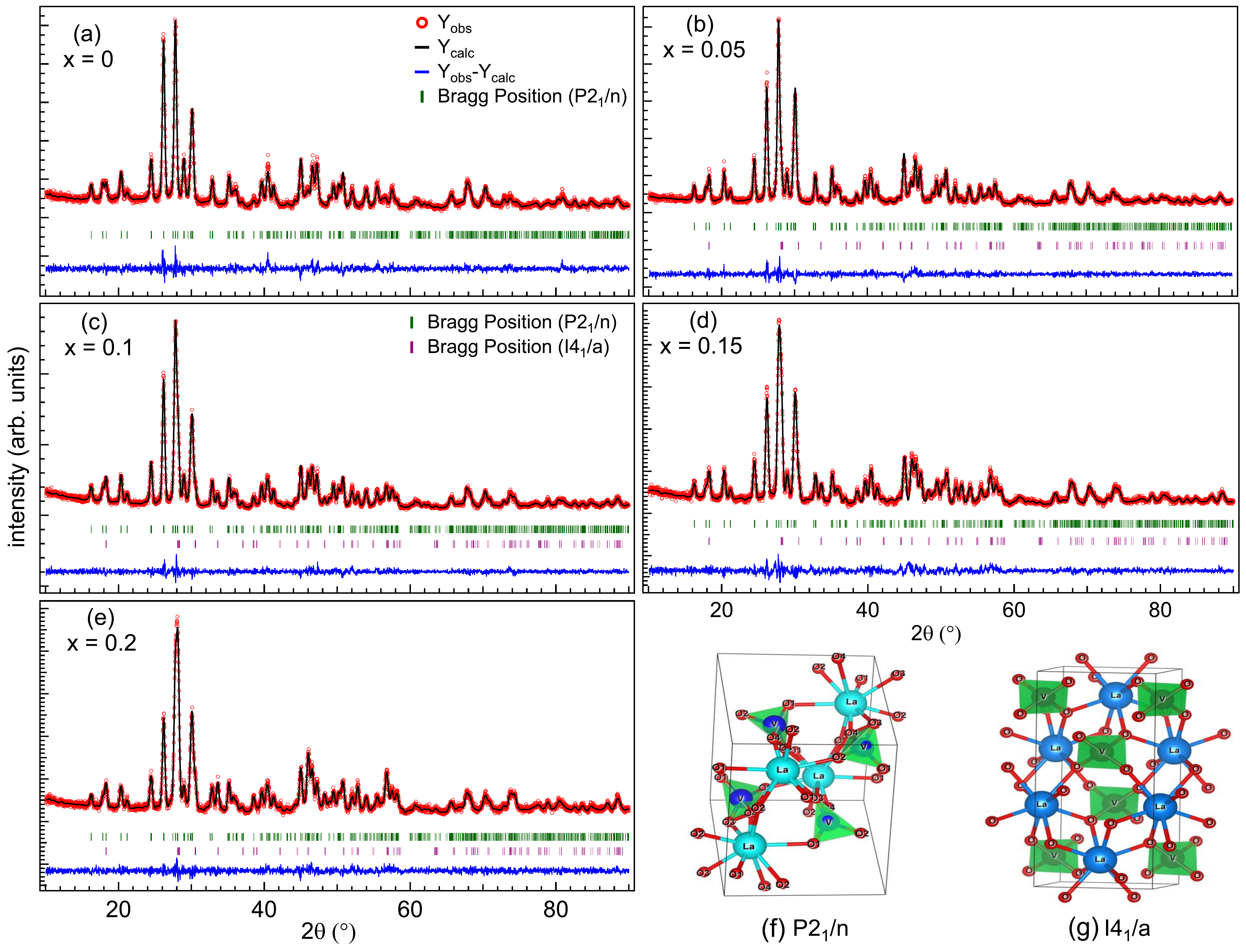}
\caption{The Rietveld refined x-ray diffraction patterns of LaV$_{1-x}$Nb$_x$O$_4$ (a) $x=$ 0, (b) $x=$ 0.05, (c) $x=$ 0.1, (d) $x=$ 0.15 and (e) $x=$ 0.2 samples. Open red circles, black solid line, and blue solid line exhibit the experimental, simulated, and the difference between experimental and simulated spectra, respectively; green and magenta vertical markers present the Bragg positions corresponding to the P2$_1$/n and I4$_1$/a space groups, respectively. (f, g) Schematic unit cell diagrams for the monoclinic monazite (P2$_1$/n) and scheelite tetragonal (I4$_1$/a) phases of LaVO$_4$ compound prepared using VESTA software \cite{MommaJAC11}, respectively, these phases are utilized to perform the Rietveld refinement of recorded x-ray diffraction patterns.}
\label{fig:XRD}
\end{figure*}

\section{\noindent ~Results and Discussion}

\begin{table*}
		\caption{The Rietveld refined lattice parameters of polycrystalline LaV$_{1-x}$Nb$_x$O$_4$ ($x=$ 0--0.2) samples with temperature-induced metastable scheelite-tetragonal phase for the $x=$ 0.1 and 0.2 samples.}
	
		\centering 
		
		\begin{tabular}{| c| c| c| c|c| c| c| c|c|}
			
			\hline
			\rule{0pt}{12pt}$~~~x~~~$ &~~~$\chi^2$~~~&~Space ~Group~&\textit{a} (\AA)& \textit{b} (\AA) & \textit{c} (\AA) &$\beta (\degree)$& ~Volume~$(\rm \AA^3)$~&~sintering~\\ 
			&&&&&&&&~temperature~\\[0.5ex]
			\hline
			
			\rule{0pt}{12pt}0 & 1.09&P2$ _1 $/n&~~7.0413(6)~~&~~7.2759(6)~~&6.7172(6) &~104.850(6)~ & 332.64(5)&1250$\rm^o$C\\ 
			\hline
			\rule{0pt}{12pt}0.05 &1.05 &~~P2$ _1 $/n - 94$\%$~~& 7.0416(3) & 7.2777(4) &6.7235(4) & 104.911(5) & 332.96(3)&1250$\rm^o$C\\
			&&I4$ _1 $/a - 6$\%$&5.3312(12)&5.3312(12)&11.712(5)&90&332.88(18)&\\
			\hline
			\rule{0pt}{12pt}0.1 &1.03 &P2$ _1 $/n - 88$\%$& 7.0409(3) & 7.2769(4) &6.7230(4) & 104.896(5) & 332.88(3)&1250$\rm^o$C\\
			&&I4$ _1 $/a - 12$\%$&5.3323(5)&5.3323(5)&~~11.7115(16)~~&90&333.00(6)&\\
			\hline	
			\rule{0pt}{12pt}0.15& 1.1&P2$_1$/n - 84$\%$&7.0431(4)  &7.2781(4)&6.7261(5)& 104.932(6)  &333.14(3) &1250$\rm^o$C\\
			&&I4$ _1 $/a - 16$\%$& 5.3313(4)& 5.3313(4)&11.7165(14)&90&333.01(5)&\\
			\hline
			\rule{0pt}{12pt}0.2&1.14 &P2$_1$/n - 72$\%$& 7.0395(4) & 7.2760(5)&6.7262(5) &104.921(6) & 332.89(4)&1250$\rm^o$C\\
			&&I4$_1$/a - 28$\%$&5.3312(3)&5.3312(3)&11.7128(11)&90&332.89(4)&\\
			\hline
			\hline
		\rule{0pt}{12pt}	0.1 & 1.44&I4$_1$/a&5.3248(2)&5.3248(2) & 11.7311(6)&90& 332.61(2)&1450$\rm^o$C\\ 
			\hline
			\rule{0pt}{12pt}0.2 & 1.47&I4$_1$/a&5.3289(2)&5.3289(2) & 11.7478(7)&90& 333.60(3)&1450$\rm^o$C\\ 
			\hline
		\end{tabular}
		\label{tab:XRD}
		
	\end{table*}

Fig.~\ref{fig:XRD} shows the Rietveld refined room temperature powder x-ray diffraction (XRD) patterns of polycrystalline LaV$_{1-x}$Nb$_x$O$_4$ ($x=$ 0--0.2) samples. We observe that the measured XRD pattern is well fitted using a monoclinic space group (P2$_1$/n) for the $x=$ 0 sample. Interestingly, the peak at 2$\rm\theta$-value of 17.83$^o$ (110) diminishes in intensity, whereas the peak at 27.77$^o$ (120) splits into two peaks with the Nb substitution. We compare the XRD patterns of all the samples in Fig.~S1(a) of ref.~\cite{SI_LaVNbO4}. Moreover, we observe that a few peaks emerge at 2$\rm\theta$-values of 28.08$^o$, 33.63$^o$, 46.08$^o$, 52.85$^o$, 56.81$^o$, and 58.25$^o$ with the Nb substitution, which corresponds to the scheelite tetragonal phase with the (112), (020), (024), (116), (312), and (224) planes, respectively \cite{DavidMRB83}. These changes in the XRD patterns are compared in Fig.~S1(a) of ref.~\cite{SI_LaVNbO4}, where the evolution of new peaks is highlighted with a black solid line. Notably, the XRD patterns for the $x>$ 0 samples can not be fitted by considering a single space group P2$_1$/n, as discussed and shown in Fig.~S1(b) of ref.~\cite{SI_LaVNbO4}. Therefore, a combination of monoclinic (P2$_1$/n) and scheelite-tetragonal space groups (I4$_1$/a, \#88) has been used to fit the XRD patterns for the $x=$ 0.05--0.2 samples. We found that the Nb substitution induced structural transformation prevail in the samples for $x\ge$ 0.05 concentration due to the larger ionic radius of Nb$^{5+}$ (0.48~\AA) ions as compared to V$^{5+}$ ions (0.355~\AA) in a tetrahedral coordination \cite{ShannonACB69}. Note that the phenomenon of structural phase transformation from monoclinic (P2$_1$/n) to tetragonal (I4$_1$/amd) phase is well investigated in LaVO$_4$ with the cationic substitution of other lanthanides (Ln$^{3+}$) via reducing the average ionic radii of the atoms present at the La site, as the structural transformation into tetragonal phase lead to the remarkable enhancement in the photoluminescence properties \cite{ParkPB10, RambabuMRB00}. Interestingly, in our case the substitution of Nb cations at V site lead to the structural phase transformation from monoclinic (P2$_1$/n) to a scheelite-tetragonal (I4$_1$/a) phase. Since the end members, LaVO$_4$ and LaNbO$_4$, exist in the monoclinic monazite (P2$_1$/n) and fergusonite monoclinic (I2/a) phase in the ambient conditions, respectively, it was expected that the cationic substitution of Nb in LaVO$_4$ will induce a structural transformation from P2$_1$/n to I2/a space group \cite{GuoIC17, RiceACB76, HuseJSSC12}. However, there are several reports on the orthovanadate compounds, which explain a pressure-induced phase transformation between the various crystal symmetries, e.g., zircon-tetragonal, scheelite-tetragonal, fergusonite-monoclinic, and BaWO$_4$-II type, etc \cite{GargJAP09, ErrandoneaPMS18, ErrandoneaPRB09, ErrandoneaJPCC16, ChengOM15}. This structural transformation from one phase to another is correlated to the occupation of $f-$electrons of the lanthanide cations and their ionic radii \cite{ErrandoneaPMS18}. 

Note that a competition between the scheelite to fergusonite and scheelite to P2$_1$/n phase transition exists in the larger R cation compounds \cite{ManjonPRB06}. However, for a complete structural transformation, i.e., in order to synthesize the pure scheelite tetragonal phase of RVO$_4$ compounds, a high hydrostatic pressure of about 25 GPa is required \cite{ErrandoneaPRB09, HuangIC12}. Here, we found a phase transformation from the fergusonite to scheelite phase during the sample synthesis and at room temperature we obtained a signature of the scheelite phase in the recorded x-ray diffraction patterns of the final product [see Fig.~\ref{fig:XRD}(b--e)]. Note that, a fergusonite to scheelite transformation is a second-order and obtained from the cyclic rotation of axes using a transformation matrix and the long-axis (unique b-axis) in fergusonite phase transforms into the c-axis of the scheelite phase. This phase transformation of the unit cell is well accompanied by the movement of the atoms (change in the Wyckoff positions), i.e., the La and V(Nb) cations move with the same magnitude, while the O atoms move in such a way that reduces the bond-length distortion index \cite{MullerOSP13, ErrandoneaJPCC16}. In short, the fergusonite phase is a compressed and less distorted version of the scheelite phase, with the $\beta$ value of 94$\rm^o$ as compared to 90$\rm^o$ for scheelite phase. There is a very effective way to visualize these structural phase transformations using group-subgroup relations by making a B$\rm\ddot{a}$rnighausen tree and applying the translationgleiche and klassengleiche transformations \cite{MullerOSP13}. Since it is well known that the t-LaVO$_4$ possess the superior optical properties as compared to the phases with monoclinic symmetry \cite{PanchalJAP11}, hence Nb substitution induced emergence of scheelite-tetragonal phase lead to an enhancement of luminescence properties in the LaV$_{1-x}$Nb$_x$O$_4$ samples. The Rietveld refined unit cell parameters are presented in table-I, where the lattice parameter values for the $x=$ 0 sample are in good agreement with the previous reports \cite{GuoIC17, RiceACB76}. Moreover, the lattice parameter values for the $x=$ 0.05 sample increases with the Nb substitution and a scheelite phase also emerges. For the samples $x >$0.05, we do not observe a systematic change in the unit cell parameters, which possibly caused by the enhancement in the proportion of the scheelite phase. Guo $et~al.$ have synthesized the $x=$ 0.1 sample to study its microwave dielectric properties and obtained a composite phase of scheelite and monoclinic; however, they did not extract the relative phase fraction and lattice parameters \cite{GuoIC17}.

\begin{figure}
\includegraphics[width=3.4in]{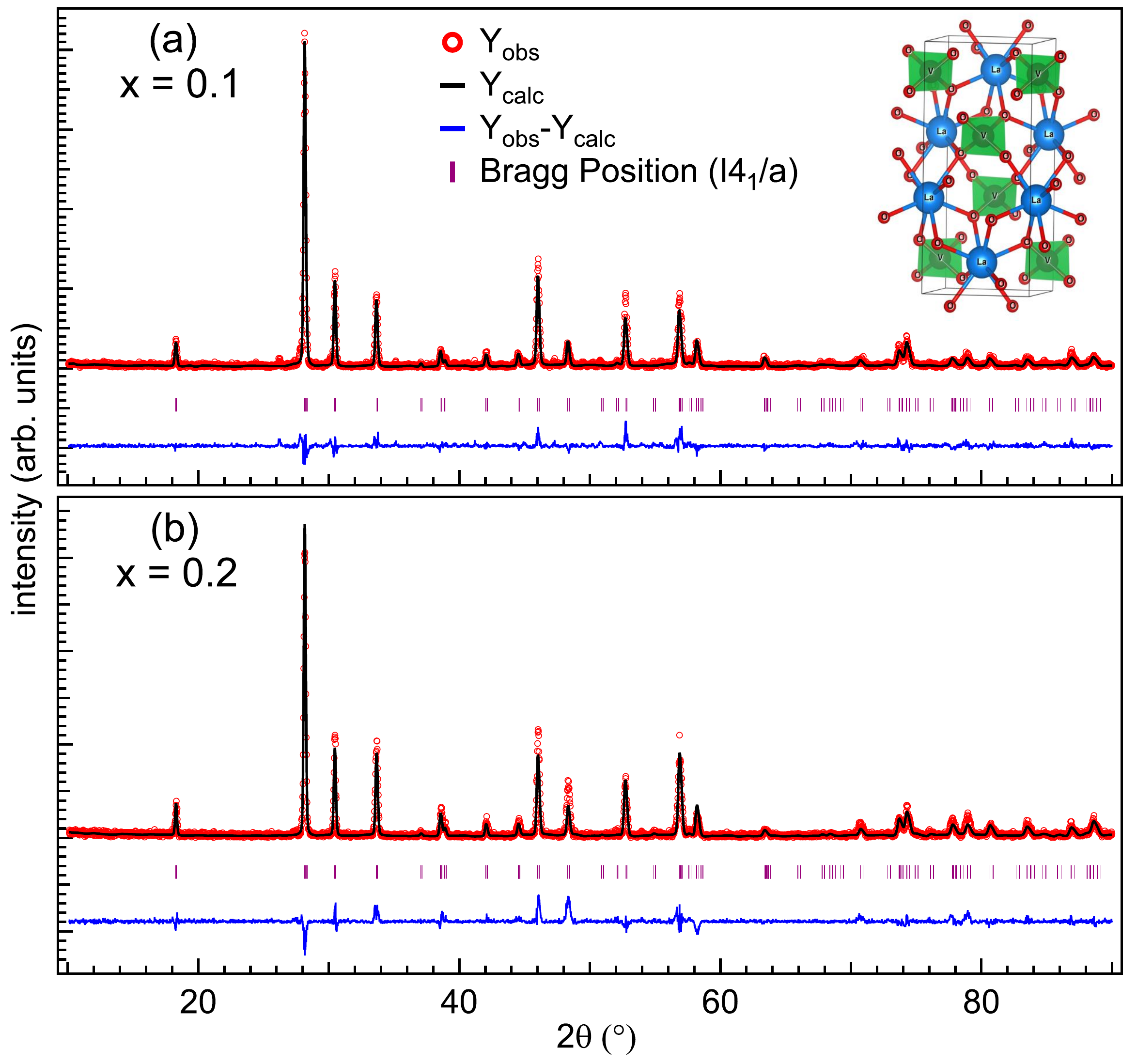}
\caption{The Rietveld refined x-ray diffraction patterns of metastable LaV$_{1-x}$Nb$_x$O$_4$ (a) $x=$ 0.1, and (b) $x=$ 0.2 samples sintered at 1450$\rm^o$C, fitted using a scheelite-tetragonal phase (space group, I4$_1$/a, \#88). Open red circles, black solid line and blue solid line exhibit the experimental, simulated, and the difference between experimental and simulated spectra, respectively. Magenta vertical markers present the Bragg positions corresponding to the I4$_1$/a space group, inset in (a) shows the schematic unit cell diagram for the scheelite-tetragonal phase utilized to fit the recorded x-ray patterns.}
\label{fig:XRD_T}
\end{figure}

It has been reported that the metastable structure of tetragonal LaVO$_4$ is commonly stabilized through a 'soft chemical' process, e.g., the hydrothermal method was used to stabilize the tetragonal phase of LaVO$_4$ \cite{OkramIC14, FanJPCB06}. Moreover, there are few studies related to the hydrostatic pressure-dependent structural phase transformation from m- to t-LaVO$_4$ phase \cite{ChengOM15, ErrandoneaJPCC16}. Interestingly, in order to obtain a stable tetragonal phase, which shows superior luminescent properties, we use high-temperature sintering of LaV$_{1-x}$Nb$_x$O$_4$ ($x=$ 0.1, 0.2) samples at 1450$\rm^o$C for 13~hrs and found a reversible transformation into a metastable scheelite-tetragonal phase [see Figs.~\ref{fig:XRD_T}(a, b)]. However, this metastable phase is again transformed to the initial stable phase (a mixture of monoclinic and scheelite-tetragonal phases). The presence of a scheelite-tetragonal phase in the $x=$ 0.1, and 0.2 samples at room temperature in a certain fraction act as a nucleation center to transform this into a complete scheelite tetragonal phase. The Rietveld refinement of this metastable phase in LaV$_{1-x}$Nb$_x$O$_4$ ($x=$ 0.1, 0.2) samples is performed using the space group I4$_1$/a ($\#$88) and shown in Figs.~\ref{fig:XRD_T}(a, b). The Rietveld refined lattice parameters for the metastable phase are also included in the table-I. We observe a monotonous increase in the lattice parameters and unit cell volume with an increase in the Nb concentration. Interestingly, this scheelite-tetragonal phase induced by the substitution of larger size Nb ions is further supported by the sintering temperature, where we observe a complete reversible structural phase transformation. Therefore, this phase transformation induced with the high-temperature sintering in the $x=$ 0.1, and 0.2 samples manifests that a scheelite-tetragonal phase can be stabilized at higher temperatures. The energy dispersive x-ray measurements performed at room temperature (not shown here) confirm the stoichiometric compositional ratio of constituent elements and their homogeneity in all the LaV$_{1-x}$Nb$_x$O$_4$ ($x=$ 0--0.2) samples.

\begin{figure*}
	\centering
	\includegraphics[width=7.1in]{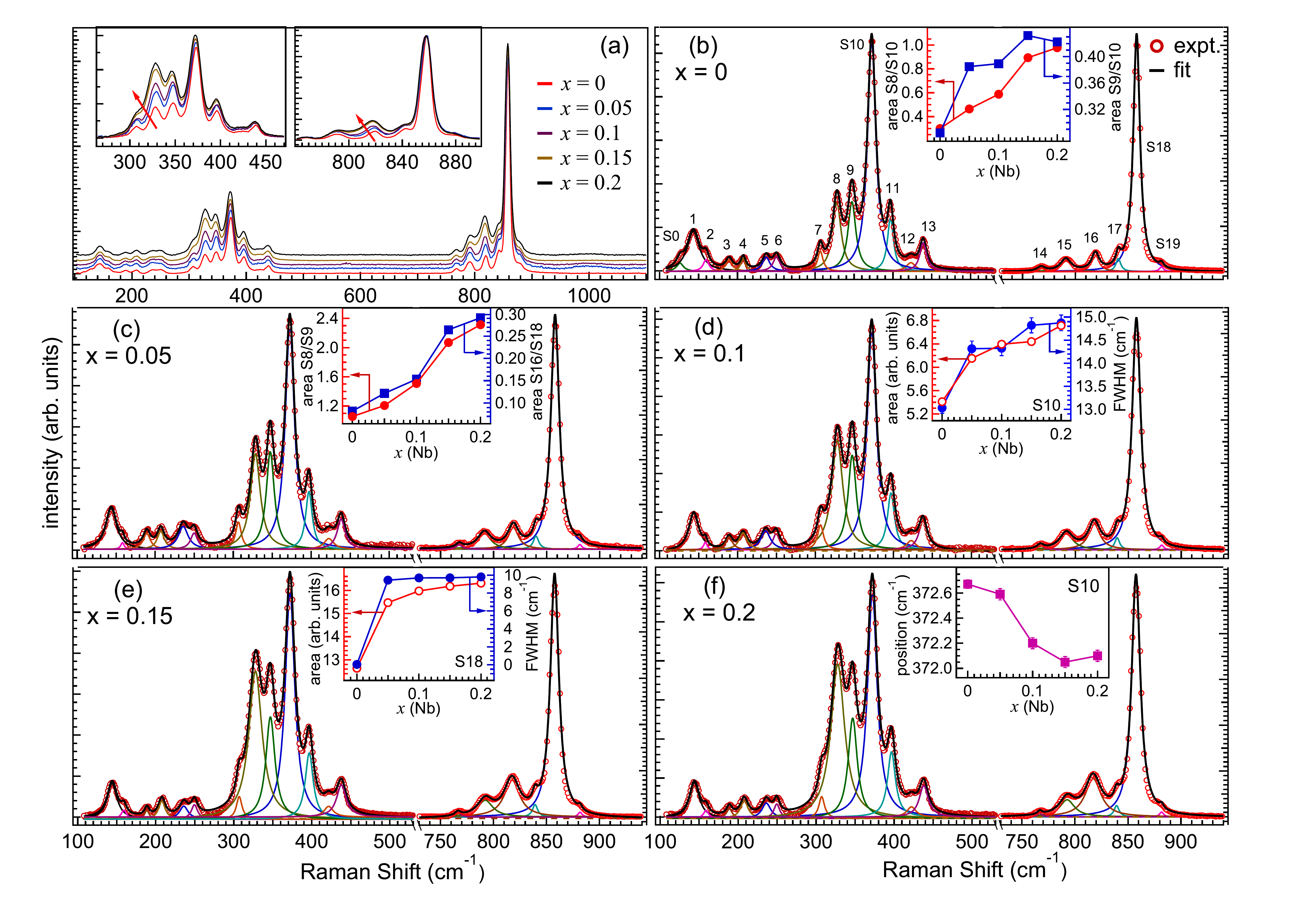}
	\caption{(a) The room temperature Raman spectra of LaV$_{1-x}$Nb$_{x}$O$_4$ ($x=$ 0--0.2) samples, left and right inset present a closer view of the as-recorded Raman modes in the lower (250-470~cm$^{-1}$) and higher (760-910~cm$^{-1}$) wavenumber regions, respectively. (b-f) Solid lines present the fitted individual modes of the recorded Raman spectra for LaV$_{1-x}$Nb$_x$O$_4$ ($x=$ 0--0.2) samples, respectively, using the Lorentzian peak function and solid thick black line presents the total fit of the measured spectra. Insets in (b) presents the variation of the relative area of S8/S10 and S9/S10 Raman modes with $x$ on the left and right-axis, respectively, (c) shows the dependence of relative area of S8/S9 and S16/S18 Raman modes with $x$ on the left and right-axis, respectively, (d, e) display the variation of the integrated area and FWHM values for the S10 (373~cm$^{-1}$) and S18 (858~cm$^{-1}$) individual Raman modes with $x$ plotted on the left and right-axis, respectively, and (f) presents the variation of the peak position of the S10 individual Raman mode with $x$, plots in the insets are marked with arrows to their respective axis.}
	\label{fig:Raman}
\end{figure*}

In Fig.~\ref{fig:Raman}(a) we present the room temperature Raman spectra of LaV$_{1-x}$Nb$_{x}$O$_4$ ($x=$ 0--0.2) samples, which are shifted vertically for a clear presentation, and the left and right insets in this graph highlight the enhancement in the Raman spectral intensity with the Nb substitution and marked with the red arrows. The recorded spectra were deconvoluted and fitted using the Lorentzian peak shape function in Figs.~\ref{fig:Raman}(b-f), with twenty individual Raman modes marked with S0--S19. It was found that 14 Raman peaks are between 100--600~cm$^{-1}$ and 6 peaks are in the range of 750--900~cm$^{-1}$, which have been listed in Table \ref{Tab:Raman}. In the LaVO$_4$ monazite structure, all the atoms possess 4$e$ Wyckoff position and the symmetry decomposition of the zone center phonons are obtained using the point group symmetry of 2/m \cite{ErrandoneaJPCC16}. According to the group theory calculations, m-LaVO$_4$ consists of 72 vibrational modes (18B$_u$ + 18A$_u$ + 18A$_g$ + 18B$_g $), which includes 3 acoustic modes (A$_u$ + 2B$_u$), 33 infrared-active modes (16B$_u$ + 17A$_u$) and 36 Raman active modes (18A$_g$ + 18B$_g$) \cite{SantosJAP07, PanchalPRB11}. Here, we are using the Mulliken symbols, where notations `A' and `B' represent that the vibrations are symmetric and antisymmetric with respect to the principal axis of symmetry, respectively. And, the subscripts `g' and `u' represent that the vibrations are symmetric and antisymmetric with respect to a center of symmetry, respectively. One can differentiate A$_g$ and B$_g$ modes experimentally using the polarized Raman measurements. For the m-LaVO$_4$ compound, there have been few reports on the Raman spectra \cite{SunJAP10, ErrandoneaJPCC16, JiaEJIC10, ChengOM15}, and we found a good agreement between our observed data and those in the references \cite{ErrandoneaJPCC16} and \cite{JiaEJIC10}. In RVO$_4$, the Raman modes can be classified as internal and external modes. The RVO$_4$ can be considered as composed of two sublattices, R and VO$_4$ units where the internal modes are due to vibration of VO$_4$ unit only, while external modes are due to the vibrations of both R and VO$_4$ units. In an ideal situation the modes due to the strong V-O bond in VO$_4^{3-}$ tetrahedron were measured in a saturated aqueous solution of Na$_3$VO$_4$, which can be characterized as $\nu_1$(A$_1$), $\nu_2$(E), $\nu_3$(F$_2$), and $\nu_4$(F$_2$). The position and other details of these Raman modes can be found in ref.~\cite{PopeSA95}. The vibrational spectra of LaVO$_4$ can be divided into three groups, (i) the low-frequency ($<$ 240~cm$^{-1}$) region, (ii) the middle-frequency (270--450~cm$^{-1}$) region, and (iii) the high-frequency region (850--970~cm$^{-1}$), which are predominantly due to the translation of La atoms, the bending vibration of O--V--O bonds, and stretching vibration of O--V--O bonds, respectively \cite{SunJAP10}. The higher frequency stretching and bending modes are due to the shortest V--O bonds. In the m-LaVO$_4$, the Raman spectra are complex due to the presence of distorted VO$_4$ tetrahedron having four different V--O bond lengths. However, some modes are related to the tetrahedral symmetry like most intense A$_g$ ($\nu_1$) breathing mode at 858~cm$^{-1}$, A$_g$ ($\nu_2$) bending mode at 373~cm$^{-1}$, and the antisymmetric B$_g$ ($\nu_4$) mode at 439~cm$^{-1}$\cite{ErrandoneaJPCC16}. Also, Errandonea \textit{et al.} found that no mode possesses pure $\nu_3$ characteristics, there exists always a mixture of $\nu_1$ and $\nu_3$ modes \cite{ErrandoneaJPCC16}. We have compared the peak positions of individual Raman modes from the reported experimental \cite{ErrandoneaJPCC16, JiaEJIC10, ChengOM15} and theoretical \cite{SunJAP10, ErrandoneaJPCC16} values in Table-II. Here our experimental values ($\omega_{obs}$) are found to be in close agreement with ref.~\cite{ErrandoneaJPCC16}; therefore used for the mode assignment.

\begin{table}
	\caption{The experimentally observed frequencies ($\omega_{obs}$) of the individual Raman modes in polycrystalline monazite LaVO$_4$ sample at room temperature with a position accuracy within $\pm$1~cm$^{-1}$. The peak positions of these modes are compared with the reported theoretical ($\omega_{th}$) values in the refs.~\cite{ErrandoneaJPCC16} \& \cite{SunJAP10} and previously obtained experimental values ($\omega_{exp}$) in the refs.~\cite{ErrandoneaJPCC16, JiaEJIC10, ChengOM15}. The values ($\omega_{obs}$) are in close agreement with ref.~\cite{ErrandoneaJPCC16}; therefore the same has been adopted for the mode assignment in the present manuscript.}
	\vskip 0.5cm
	\centering 	
	\begin{tabular}{|c|c|c| c|c|c|c|c|}
		\hline 
		peak & $\omega_{obs}$ & $\omega_{th}$ \cite{ErrandoneaJPCC16} & $\omega_{exp}$ \cite{ErrandoneaJPCC16}  & $\omega_{th}$ \cite{SunJAP10} & $\omega_{exp}$\cite{JiaEJIC10} & $\omega_{exp}$ \cite{ChengOM15} \\ 
		&cm$^{-1}$&cm$^{-1}$ &cm$^{-1}$&cm$^{-1}$&cm$^{-1}$&cm$^{-1}$\\[0.5ex]
		\hline
		S0& B$_g$(126) & B$_g$(127) & 127 & A$_g$(126) &127 & 124.2 \\
		S1& A$_g$(143) & A$_g$(143) & 146 & B$_g$(143) &147 & 143.8 \\
		S2& B$_g$(160) & B$_g$(158) & 160 & B$_g$(170) &158 & 156.6 \\
		S3& A$_g$(189) & A$_g$(188) & 189 & --  & 189 & 187.3 \\
		S4& B$_g$(208) & B$_g$(204) & 209 & A$_g$(203) & 208 & 204.7 \\
		S5& A$_g$(237) & A$_g$(230) & 235 & A$_g$(232) & 238 & 242.1 \\
		S6& A$_g$(250) & A$_g$(252) & 252 & -- & 251 & 260.4 \\
		S7& B$_g$(307) & B$_g$(316) & 309 & B$_g$(315) & 309 & 306.2\\
		S8& A$_g$(328) & A$_g$(336) & 326 & A$_g$(334) & 329 & 326.3\\
		S9& A$_g$(347) & A$_g$(355) & 349 & -- & 349 & 345.8\\
		S10& A$_g$(373) & A$_g$(380) & 373 & B$_g$(378) & 374 & 370.7\\
		S11& B$_g$(397) & B$_g$(389) & 397 & B$_g$(394) & 398 & 394.8\\
		S12& A$_g$(423) & A$_g$(423) & 426 & -- &-- & 420.7\\
		S13& B$_g$(439) & B$_g$(427) & 439 & -- & 440 & 436.4\\
		S14& A$_g$(768) & A$_g$(784) & 768 & -- & 770 & 766.5\\
		S15& B$_g$(791) & B$_g$(799) & 790 & -- & 794 & 792.1\\
		S16& A$_g$(819) & A$_g$(806) & 819 & -- & 819 & 817.5\\
		S17& A$_g$(841) & A$_g$(836) & 843 & -- &-- & 840.9\\
		S18& B$_g$(858) & B$_g$(861) & 855 & A$_g$(865) & 859 & 856.5\\
		S19& B$_g$(881) & B$_g$(892) & 882 & A$_g$(883) & -- & -- \\
		\hline
	\end{tabular}
	\label{Tab:Raman}
\end{table}

Further, we have analyzed the variation in the integrated area of the selected Raman modes and their full-width-at-half-maximum (FWHM) with the Nb concentration ($x=$ 0--0.2). We observe that the S0 (A$_g$) individual mode (126~cm$^{-1}$) is only present for the $x=$ 0 sample and completely disappears for the Nb substituted samples ($x >$0). This Raman mode is attributed to the translational of the La atoms, which indicates that the emergence of an additional phase with Nb substitution diminishes the intensity of this mode. Moreover, we found that the major peaks, which have the significant intensity contribution in the Raman spectra, i.e., S8, S9, S10, S15, S16, and S18 exhibit a monotonous enhancement in the integrated intensity as well as in the full-width-at-half-maximum (FWHM) with the increase of Nb concentration. This behavior is related to the fact that the increase in the scheelite-tetragonal phase with Nb substitution is responsible for the enhancement in the Raman modes of the Nb substituted samples. These monotonous effects on the vibrational spectra are correlated to the substitution induced deformation of the VO$_4^{3-}$ tetrahedra and also can be seen in the FTIR spectra of the Nb substituted samples, discussed later. To see the variation in the intensity of S8 and S9 Raman modes with $x$, we plot their relative integrated area ratios as compared to the intense S10 mode on the left and right axis of the inset in Fig.~\ref{fig:Raman}(b). We found an enhancement in the relative area ratio of S8/S10 and S9/S10 modes with increasing $x$, see the inset in Fig.~\ref{fig:Raman}(b). It can be seen that the intensity of the S8 mode at 328~cm$^{-1}$, which is lower as compared to the S9 mode for the $x=$ 0 sample, gradually increases with $x$ and becomes higher for the $x=$ 0.2 sample [see Figs.~\ref{fig:Raman}(b-f)]. Therefore, we present the variation of the relative integrated area ratio of the Raman modes S8 and S9 on the left axis of the inset in Fig.~\ref{fig:Raman}(c), which was found to increase with the $x$. Similarly, the variation of the relative area ratio of S16 Raman mode with respect to the most-intense S18 mode is shown with $x$ on the right axis of the inset in Fig.~\ref{fig:Raman}(c), which exhibits an increasing trend. These results indicate that the Nb substitution-induced emergence of the scheelite-tetragonal phase enhances the intensity of the selected Raman modes significantly with $x$. Further, we analyze the behavior of the most intense Raman modes in the lower and higher wavenumber region, i.e., S10 (373~cm$^{-1}$) and S18 (858~cm$^{-1}$). A variation of the integrated area and FWHM values of S10 (A$_g$) mode with $x$ is shown on the left and right-axis, respectively [see inset of Fig.~\ref{fig:Raman}(d)]. We observe that the integrated area and FWHM values of A$_g$ bending mode (S10) increase monotonically with the $x$. Also, we present the behavior of the peak position of S10 mode [see inset of Fig.~\ref{fig:Raman}(f)] with $x$ and found a monotonous decrease in the peak position, which indicates that the A$_g$ bending mode frequency decreases with deformation of VO$_4$ tetrahedra owing to the Nb substitution. Interestingly, for the S18 Raman mode (B$_g$) we found that the area and FWHM values exhibit a jump in the increment from the $x=$ 0 to 0.05 sample, and then a monotonous increase in the values with the Nb concentration, see Fig.~3(e). However, we found that the position of the S18 mode does not change despite the Nb substitution in the system (not shown), i.e., breathing vibrations associated with this mode are unaffected due to tetrahedral distortions. In addition to that, we have also investigated the peak position of the other fitted components with the Nb substitution, and found that there is no significant shift in the peak position of the S7, S8, S11, S12, S13, S15, and S20 Raman modes, whereas S9, S16, and S17 (all A$_g$) Raman modes exhibit a monotonous change with $x$ (not shown here). The most intense S10 (A$_g$) and S18 (B$_g$) Raman modes manifest a remarkable change in the integrated area/FWHM with the Nb substitution in the VO$_4$ tetrahedra. These results further motivate us to investigate the local structure of these samples using x-ray absorption spectroscopy (XAS). 

\begin{figure}[h]
\includegraphics[width=3.55in]{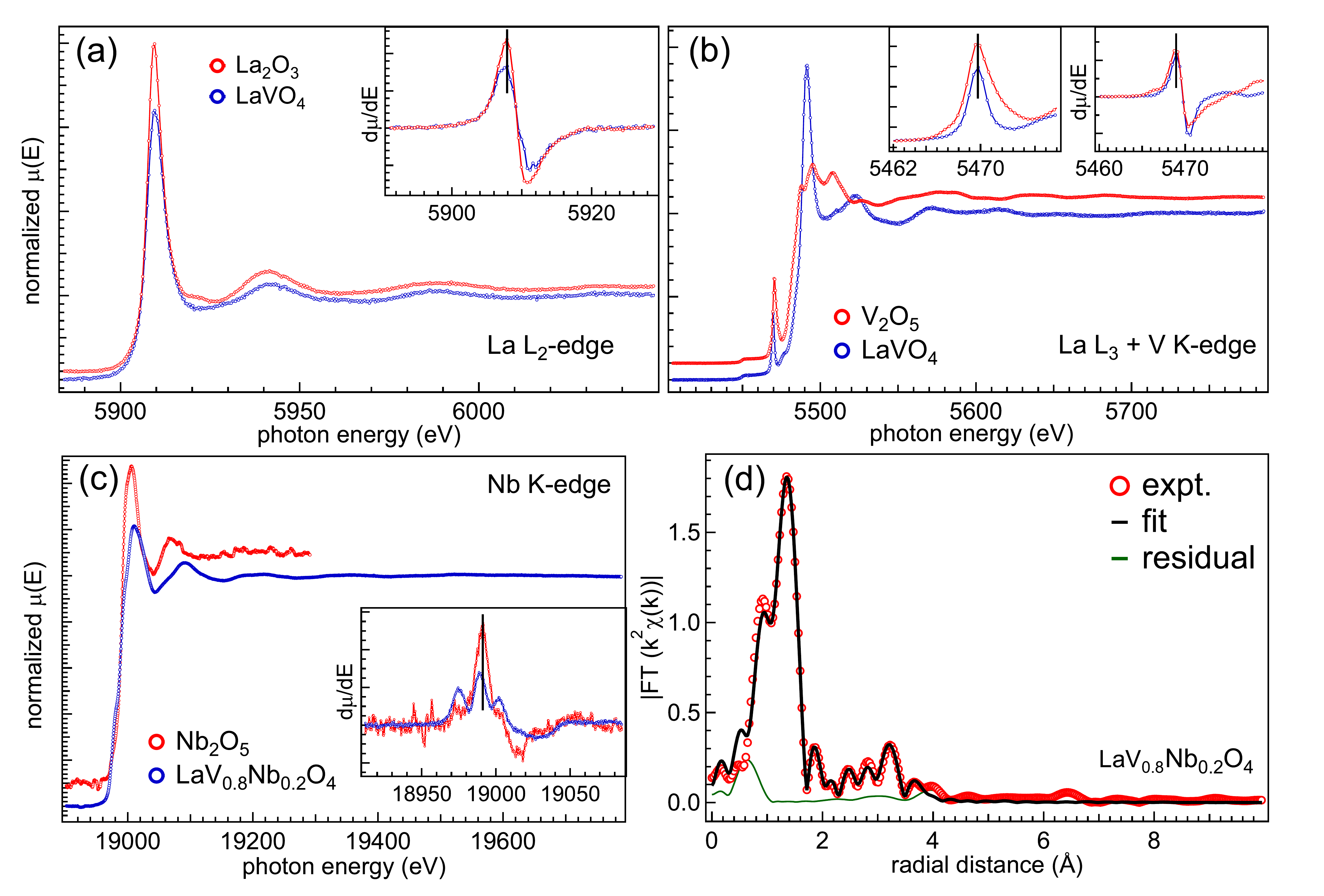}
\caption{The room temperature x-ray absorption spectra recorded in the transmission mode for, (a) LaVO$_4$ and reference La$_2$O$_3$ samples at La L$_2$-edge, inset shows a comparison of the peak position in the first-order derivatives, (b) the combined La L$_3$ and V K-edges of LaVO$_4$ and V$_2$O$_5$ samples where left inset highlights the pre-edge peak and the right inset exhibit the comparison of maximum obtained in the first-order derivative, (c) the Nb K-edge of LaV$_{0.8}$Nb$_{0.2}$O$_4$ and Nb$_2$O$_5$ samples where the inset shows a closer view of the maxima in the first-order derivative, (d) the Fourier transform of the EXAFS spectrum at the Nb K-edge along with curve fitting for the LaV$_{0.8}$Nb$_{0.2}$O$_4$ sample. Note that the reference spectrum in the main panels (a-c) is vertically shifted by a constant factor of 0.1 for clear presentation.}
\label{fig:XAS}
\end{figure}

Therefore, the XAS measurements were performed for the La L$_2$-, Nb K-, and La L$_3$+V K-edges at room temperature for the $x=$ 0 and 0.2 samples. We have calibrated each spectrum with the reference metal foils utilizing the maxima in the first-order derivative about the inflection near the edge-jump. The normalized x-ray absorption spectra for LaVO$_4$ and reference La$_2$O$_3$ samples measured at the La L$_2$-edge (5891~eV) are presented in Fig.~\ref{fig:XAS}(a), where the inset shows the derivative about the first inflection near the edge jump and a solid line along the maxima (5907.9~eV) affirm that La cations exist in the trivalent oxidation state analogous to the La$_2$O$_3$ sample. As the photon energies of the V K-edge (5465~eV) and La L$_3$-edge (5483~eV) are close to each other, an absorption measurement at one of them will eventually result in the strongly overlapped absorption spectra corresponding to both the edges for the LaVO$_4$ sample. In Fig.~\ref{fig:XAS}(b) we present the normalized absorption spectra at these (V K- and La L$_3$-) edges for the LaVO$_4$ sample as well as for V$_2$O$_5$ as reference. A pre-edge peak (near 5469.8~eV) in Fig.~\ref{fig:XAS}(b) appears due to the tetrahedral coordination of V ions in the monoclinic structure and confirms the $d\rm^0$ configuration consistent with the earlier reports, as the intensity of the pre-edge feature is highest for the $d\rm^0$ compounds and monotonously decreases to zero for $d\rm^{10}$ configuration \cite{ChandraActa20, YamamatoXRS08}. This pre-edge peak in the V K-edge emerges due to the electric dipolar transition to the $p$ component in the $p-d$ hybridized orbitals and some smaller contributions arise from the quadrupolar transitions within the same orbitals \cite{YamamatoXRS08}. Note that this pre-edge feature is solely emerging due to the V K-edge and unaffected by the contributions from the La L$_3$-edge. A comparison of the measured spectra with reference V$_2$O$_5$ sample at the pre-edge peak (near 5469~eV) in the left inset and the first-order derivative in the right inset of Fig.~\ref{fig:XAS}(b) portray a 5+ oxidation state of V ions similar to the reference V$_2$O$_5$ sample. Further, we have recorded the absorption spectrum of the Nb K-edge (18986~eV) for the LaV$_{0.8}$Nb$_{0.2}$O$_4$ sample having the highest Nb concentration and compared the maxima in the first-order derivative (around 18991~eV) with the reference Nb$_2$O$_5$ sample in Fig.~\ref{fig:XAS}(c), which reveals that the Nb ions in our sample exist in the 5+ oxidation state. Moreover, we have performed the curve fitting of the Nb K-edge extended x-ray absorption fine structure (EXAFS) for the $x=$ 0.2 sample using the Artemis program \cite{RavelJSR05} up to the radial distance of 4.1~\AA, as shown in Fig.~\ref{fig:XAS}(d). Note that, the fitted spectrum is not corrected with back-scattered and central phase shifts. In the curve-fitting procedure the atomic scattering paths were generated using the FEFF software with the help of results obtained from the Rietveld refinement of XRD pattern for the $x=$ 0.2 sample, which shows both monoclinic (72\%) and scheelite-tetragonal (28\%) phases. Our analysis manifest that in the monoclinic phase there are four unequal V/Nb--O bond lengths available in a group of two sets, i.e., 3$\times$1.804~\AA~and 1$\times$1.749~\AA, while for the scheelite-tetragonal phase we obtained that the tetrahedra consist of four equal bonds with the values 4$\times$1.842~\AA~and 4$\times$2.491~\AA. These values of bond-lengths appear due to the cumulative contribution in the atomic absorption process from the oxygen atoms present in the vicinity of the absorbing Nb atoms.

\begin{figure}
	\includegraphics[width=3.55in]{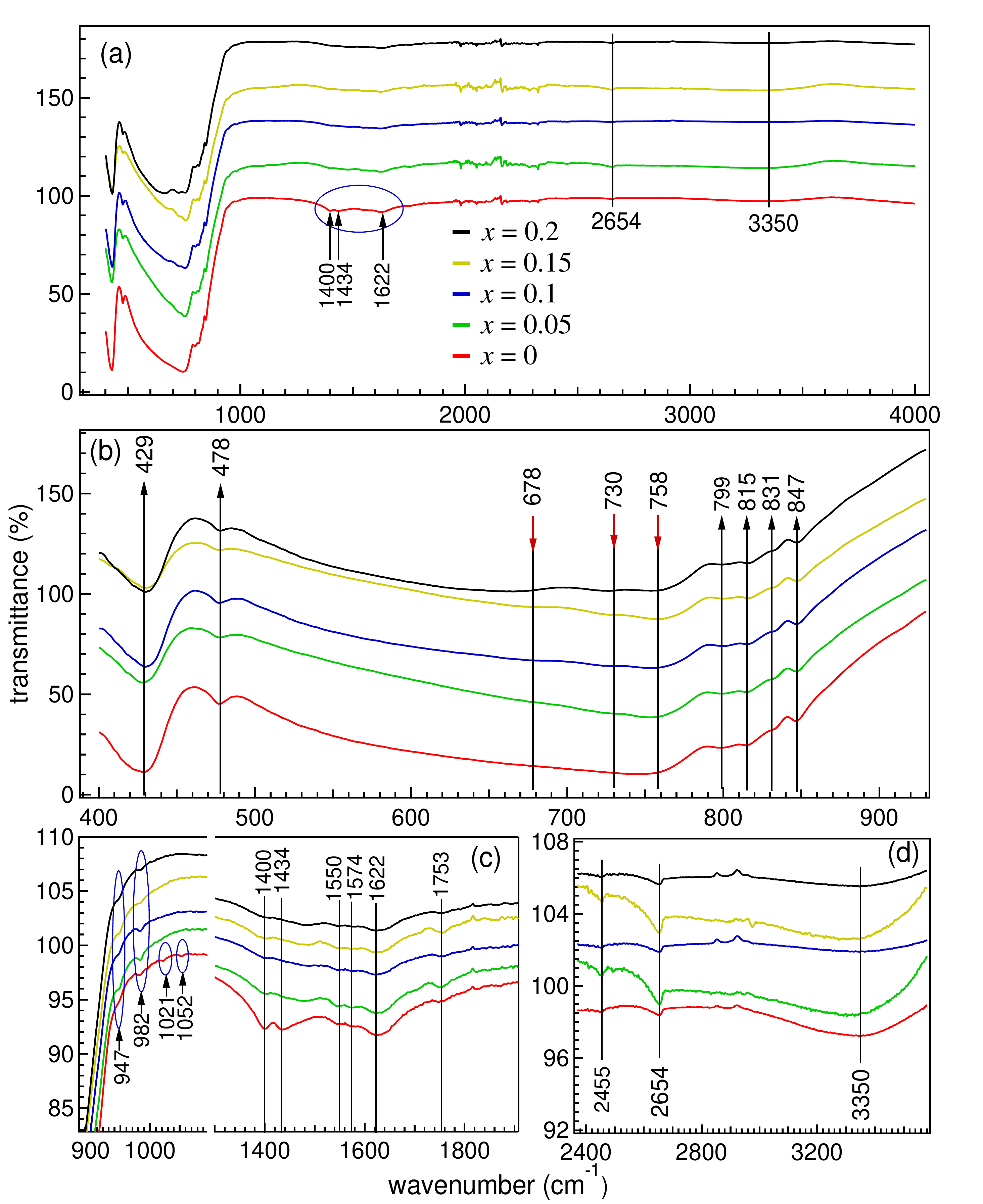}
	\caption{(a) The Fourier transform infrared (FTIR) spectra of LaV$_{1-x}$Nb$_{x}$O$_4$ ($x=$ 0--0.2) samples measured at room temperature, (b) highlights the range between 390--930~cm$^{-1}$, where black arrows mark IR modes observed for all samples and new modes due to Nb substitution are marked with the red arrows, (c, d) the IR modes between 880--1908~cm$^{-1}$ and 2370--3580~cm$^{-1}$, respectively. The spectra have been shifted vertically for clear presentation.}
	\label{fig:FTIR}
\end{figure}

Now, in order to validate the structure and composition, as well as to look for the possibility of infrared active modes, we perform the Fourier-transform infrared (FTIR) measurements of all the samples, as shown in Fig.~\ref{fig:FTIR}. The as-recorded FTIR spectra in Fig.~\ref{fig:FTIR} are vertically shifted for clear presentation, while closer views are presented in Figs.~\ref{fig:FTIR}(b--d). All the well-resolved individual peaks are marked with the black vertical arrows, whereas the additional peaks emerging with Nb substitution are marked with the red arrows [see Fig.~\ref{fig:FTIR}(b)]. The obtained spectra are similar to the other orthovanadate compounds, i.e., RVO$_4$ (R = Y, Ce-Yb) \cite{LiuAM07} and also with the compounds which possess the isolated tetrahedral VO$_4^{3-}$ groups, e.g., multi-metal orthovanadates \cite{RghuouiJMES12, FangCL99}. However, the FTIR spectra in compounds are modified due to variation in the site symmetry of VO$_4^{3-}$ anion as compared to the free-ions and the parent sample. As discussed above, the monoclinic LaVO$_4$ crystallizes into P2$_1$/n (P121/n1) space group, which manifests a primitive unit cell associated with the 2-fold screw axis along the unique axis $b$, and a glide plane $n$ perpendicular to the $b$ direction. The point group symmetry associated with the P2$_1$/n space group is 2/m \cite{ErrandoneaJPCC16}, which can be visualized as a mirror plane perpendicular to the 2-fold rotation axis. Therefore, the VO$_4^{3-}$ anion groups can be visualized at these point group symmetry positions in the monoclinic structure of LaVO$_4$ [see Fig.~1(f)]. In this regard, a detailed description of the point group symmetries for the different crystal systems of tetrahedral XO$_4^{3-}$ ions can be found in ref.~\cite{HezelSA66}. 

Moreover, the stretching ($\nu_1$ \& $\nu_3$) and bending ($\nu_2$ \& $\nu_4$) vibrations in the VO$_4^{3-}$ tetrahedra are associated with the O-V-O bonds, which exhibit the characteristic IR peaks for the different orthovanadate compounds in the range of 700-1100~cm$^{-1}$ and 400-700~cm$^{-1}$, respectively. In the recorded spectra of all the samples, we observe the well-resolved strong and sharp bands at 429, 478 cm$^{-1}$, which correspond to the $\nu_4$ bending vibrations of VO$_4^{3-}$ tetrahedra \cite{SunJAP10, ChukovaNRL17}. Also, the wide bands between 700-1100~cm$^{-1}$ are clearly observed, where a peak at ~799~cm$^{-1}$ quantifies the anti-symmetric stretching along with the other overlapped peaks near 815, 831, 847~cm$^{-1}$ correspond to the $\nu_3$ stretching vibrations of the VO$_4^{3-}$ tetrahedra \cite{SunJAP10, FangCL99, ChukovaNRL17}, as marked by arrows in Fig.~5(b). In Fig.~\ref{fig:FTIR}(c) we indicate four minima at positions 947, 982, 1021, and 1052 cm$^{-1}$ in which modes 947 and 982~cm$^{-1}$ appear due to the stretching vibrations of O-V-O bonds \cite{SunJAP10, LiuMCP09}, and the other small modes near 1021 and 1052~cm$^{-1}$ in the $x=$ 0 sample appear due to the surface impurities from V$_2$O$_5$ \cite{FangCL99} and C-O vibration \cite{OkramMSEB13}, respectively. Further, the modes at wavenumbers 1400, 1434, 1550, 1574, 1622, 1753, 2455, 2654, and 3350~cm$^{-1}$ are observed in FTIR spectra  and marked by solid black lines in Figs.~\ref{fig:FTIR}(c, d). Here, the 1400 and 1434~cm$^{-1}$ peaks appear due to the symmetric and antisymmetric vibrations of carboxylate groups, which emerge due to the reactive surface of LaVO$_4$ with CO$_2$ in air to form carbonate species, and the peaks at 1550 and 1574~cm$^{-1}$ correspond to the asymmetrical bond-stretching of carboxylate bidentate \cite{GonzalezJN21}. A strong band at 1622~cm$^{-1}$ corresponds to the O-H bending vibrations from adsorbed water molecules \cite{GonzalezJN21, OkramMSEB13}. A small peak at 1753~cm$^{-1}$ is related to the C=O stretching vibrations \cite{HadjiivanovCR21}, and the mode near 2455~cm$^{-1}$ is generally associated with the vibrations from the hydrides \cite{CoatesWiley06}. The modes near 2654~cm$^{-1}$ are related to the CH$_2$ stretching vibration and the broad minimum near 3350~cm$^{-1}$ is attributed to the O--H stretching arising from the water absorption \cite{OkramMSEB13, GonzalezJN21}. We can see that these bands (between 1400 and 3350~cm$^{-1}$) persist in all the samples, since our samples are sintered at such a high temperature, i.e., 1250$\rm^o$C, therefore these modes are expected to be associated with the surface absorbed impurities of carbon and water. It is a little difficult to assign each spectral component to a specific type of vibrational mode and especially in the stretching range of vibrations. In the group theory calculations for the vibrational properties, the origin of the IR modes is associated with the crystal symmetry of the LaVO$_4$ compound. So, due to the lower symmetry of the m-LaVO$_4$ phase (2/m), total 33 IR active modes are predicted, where strongly overlapped IR active modes appear in the form of a wideband \cite{SunJAP10}. The addition of Nb cations will distort the VO$_4$ tetrahedra and that results in the observed changes in the IR spectra of substituted samples. In the present case, all the substituted samples are in a mixed phase of monoclinic and scheelite-tetragonal in a certain proportion, as summarized in Table~I. Here, we found the emergence of three new peaks for the Nb substituted samples at 678, 730, and 758~cm$^{-1}$, which were absent for the $x=$ 0 sample, these features are marked by red arrows in Fig.~\ref{fig:FTIR}(b). These additional peaks in the IR spectra appear due to the additional symmetry modes of the scheelite-tetragonal phase in the Nb substituted samples and have their contribution from the LaNbO$_4$ phase \cite{TsunekawaMRB77}. The intensity of these peaks enhance monotonically with the Nb substitution. However, the IR mode at 478~cm$^{-1}$ diminish in intensity with the Nb substitution. An overall change in the FTIR spectra including appearance of additional peaks [highlighted by red arrows in Fig.~5(b)], and variation in the intensity of modes at 478 and near 800~cm$^{-1}$ are attributed to the effect of Nb substitution at the V site in LaVO$_4$ sample. 

\begin{figure}[h]
\includegraphics[width=3.55in]{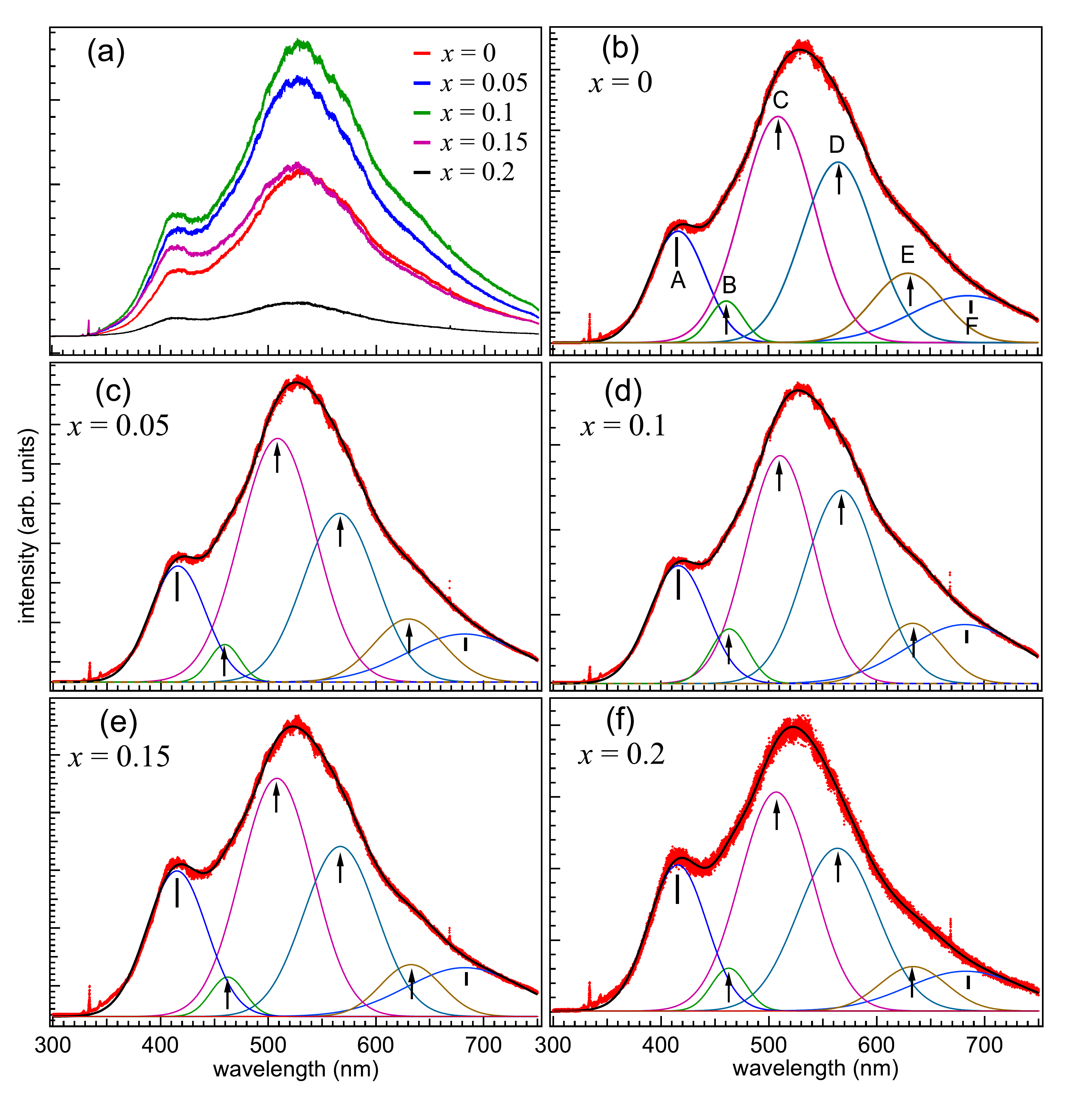}
\caption{(a) A comparison of the photoluminescence (PL) spectra of LaV$_{1-x}$Nb$_{x}$O$_4$ ($x=$ 0--0.2) samples measured at room temperature using 325~nm excitation wavelength, (b--f) deconvoluted and fitted PL spectra with six Gaussian peaks (A--F) for the $x=$ 0--0.2 samples, respectively.}
\label{fig:PL}
\end{figure}

In order to understand the photoluminescence (PL) behavior, the room temperature PL spectra of LaV$_{1-x}$Nb$_x$O$_4$ ($x=$ 0--0.2) samples were recorded in a spectral range of 300-750~nm using an excitation wavelength of $\lambda_{ex}$ = 325~nm [see Fig.~\ref{fig:PL}(a)]. We observe a broad asymmetric and strongly overlapped peak centered near 540~nm with one small peak at a lower wavelength side near 415~nm [see Fig.~\ref{fig:PL}]. The strong emission in the PL spectra manifest the rapid recombination of electron-hole pairs \cite{HeJMCA11}. The observed spectra are complex in nature having several strongly overlapped components, which emerge due to the transitions from a more complicated energy scheme. The observed spectrum for the $x =$ 0 sample is analogous to the earlier reports, where several groups have observed the strongly overlapped bands centered near 540~nm \cite{ChukovaNRL17, GulOAM19, ChukovaSSP15}; however, a small kink near 415~nm is not observed in the literature. The broad peak shape is related to the several $d$-$f$ transitions of the lanthanum orthovanadate compound \cite{ZhengaMCP15}. We have deconvoluted and fitted the measured PL spectra using six Gaussian peaks, as marked by alphabets A--F and shown in Figs.~\ref{fig:PL}(b--f) for the $x=$ 0--0.2 samples, respectively. The position of four peaks (B--E, marked by arrows) are taken from the related reference from the literature \cite{ChukovaSSP15, GulOAM19, KrumpelJPCM09}, while two peaks (A and F, marked by vertical lines) in the respective ends are inserted to fit the data with the minimum number of Gaussian peaks. The presence of six peaks in the deconvoluted PL spectra indicate the availability of more than one recombination centers for the electron-hole pairs, as the emission and excitation in these samples are all governed by the VO$_4^{3-}$ tetrahedron having the T$\rm_d$ symmetry. Interestingly, the V$^{5+}$ molecular orbitals consist of a ground state $^1A_1$ and four excited states named as $^1T_1$, $^1T_2$, $^3T_1$, and $^3T_2$ \cite{NakajimaJPCC10}. Among these states, the transitions related to the absorption process, i.e., $^1T_1$, $^1T_2$ $\leftarrow$ $^1A_1$ are allowed, whereas the emission process transitions, i.e., $^3T_1$, $^3T_2$ $\rightarrow$ $^1A_1$ are forbidden for the ideal T$\rm_d$ symmetry in terms of the spin-selection rule ($\Delta$S = 0). However, in the experimental observation, we found that these forbidden transitions become partially allowed due to the distortion of VO$_4^{3-}$ from the ideal state, which in turn changes the strength of spin-orbit coupling as it depends on the central-field potential of atoms as well as their spin and orbital angular momenta \cite{NakajimaJPCC10, OjambatiACSP20}. Interestingly, the tetrahedral distortion in VO$_4^{3-}$ caused due to cationic substitution enhances the spin-orbit interaction and hence the intensity of the forbidden transitions is observed significantly higher in these compounds. These transitions are related to bands observed near the 570 and 635~nm \cite{ChukovaNRL17}. Moreover, in m-LaVO$_4$ two bands at 550--650~nm and 650--700~nm are attributed to the $^5D_0$$\rightarrow$$^7F_1$ and $^5D_0$$\rightarrow$$^7F_2$ transitions, respectively \cite{YeJACS14}. The first is broadband which displays transitions of excited 5$d$ states to the $^2F_{5/2}$ ground state, and last is small band related to $^5D_0$$\rightarrow$$^7F_4$ transition \cite{YeJACS14, ZhengaMCP15}. The peak near 460~nm is related to the electron transitions between $^1T_1$$\rightarrow$$^1A_1$ states \cite{ChukovaSSP15}. We observe that for the $x=$ 0.1 sample, the intensity of the emission spectra increases and FWHM decreases, indicating the reduction in the recombination centers. However, for the $x=$ 0.2 sample, the PL intensity decreases and FWHM increases, which suggests the increase in the recombination centers for the electron-hole pairs. We found that the peak near 415~nm corresponds to the 2.98~eV of photon energy is related to the electron transition in V$_2$O$_5$ \cite{LeRSC18}. 

\begin{figure}[h]
	\centering
	\includegraphics[width=3.4in]{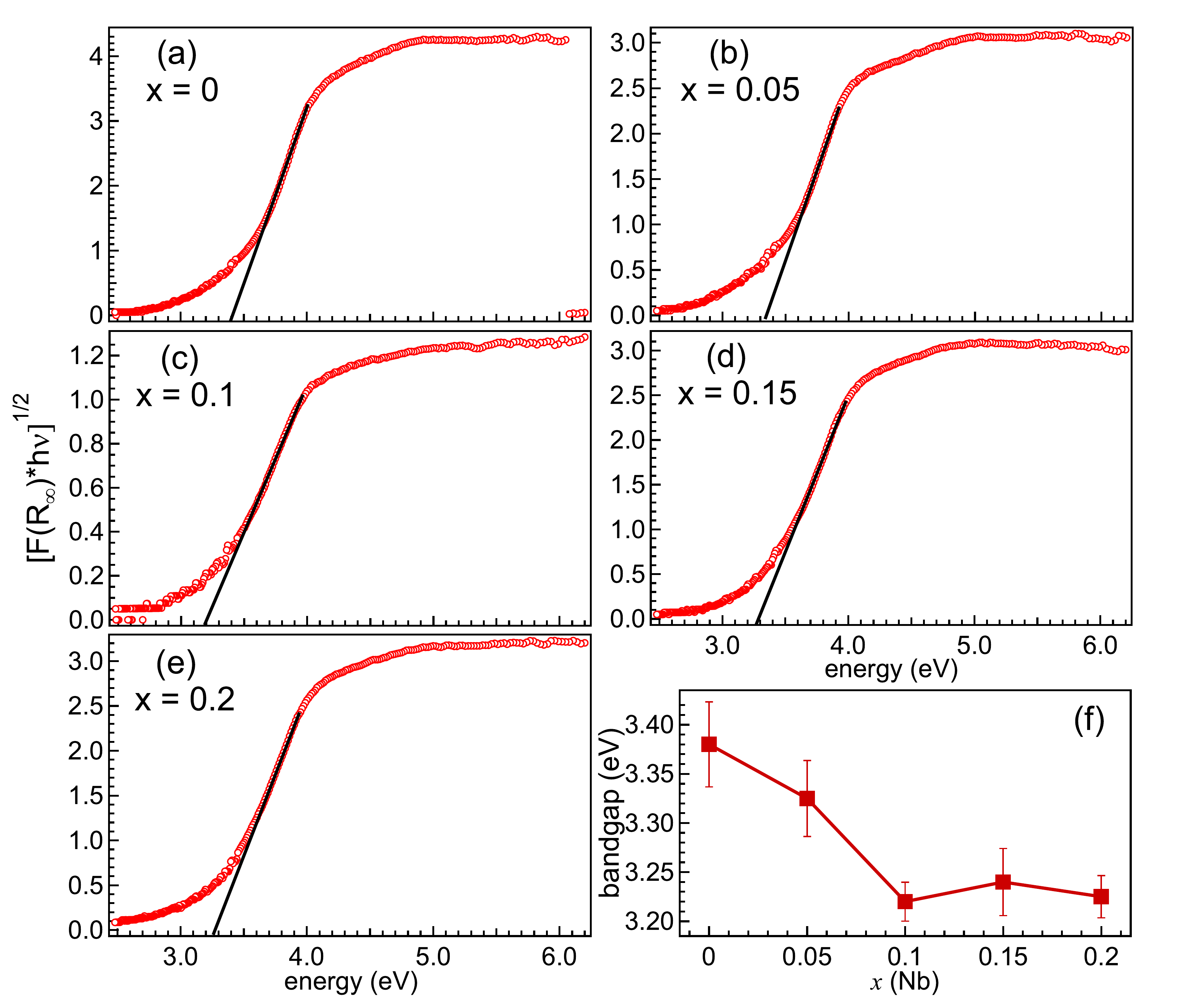}
	\caption{(a-e) The Kubelka-Munk plots for the polycrystalline LaV$_{1-x}$Nb$_{x}$O$_4$ ($x=$ 0--0.2) samples obtained from the diffused reflectance spectroscopy data, and (f) the optical bandgap as a function of $x$ calculated from the x-axis intercept along the optical absorption edge in the Tauc plot.}
	\label{fig:DRS}
\end{figure}

Finally, in order to find the bandgap we perform the diffuse reflectance measurements of all the prepared LaV$_{1-x}$Nb$_{x}$O$_4$ ($x=$ 0--0.2) samples sintered at 1250$\rm^o$C. The recorded diffuse reflectance spectra were utilized to estimate the Kubelka-Munk (K-M) function [F(R$_{\infty}$)], which can be defined in terms of diffuse semi-infinite reflectance, i.e., R$_{\infty}$=$R_{\rm sample}/R_{\rm standard}$ as \cite{KubelkaZTP31}, $$F(R_{\infty})=(1-R_{\infty})^2/(2R_{\infty})$$ Here, we have used TiO$_2$ as a standard sample in our measurements. The bandgap of samples can be estimated from the absorption threshold energy (given by the x-axis intercept) in the Tauc plot, which can be written as $$(\alpha h\nu) = k(h\nu - E_g)^{\eta}$$ where $\alpha$, $\nu$, $k$, E$_g$, and $\eta$ are the absorption coefficient, frequency, absorption constant, bandgap energy, and exponent of the equation \cite{TaucPSSB66}. The value of this exponent $\eta$ depends upon the type of the transition/bandgap in the material and can have values of 1/2, 2, 3/2, and 3 for the direct allowed, indirect allowed, direct forbidden, and indirect forbidden transitions, respectively \cite{TaucPSSB66}. In the case of diffuse reflectance spectra the $\alpha$ in Tauc plot can be replaced by the K-M function to estimate the bandgap of the material \cite{MakulaJPCL18}. Since, our material has an indirect bandgap so the value of $\eta$ is 2 and therefore we have plotted the [F(R$_{\infty}$)$\times$h$\nu$]$^{1/2}$ versus E for each sample in Figs.~\ref{fig:DRS}(a-e). The x-axis intercept of the linear fitting [presented as a solid black line in each graph of Figs.~\ref{fig:DRS}(a-e)] along the optical absorption edge for samples LaV$_{1-x}$Nb$_x$O$_4$ ($x=$ 0 -- 0.2) gives an estimation of optical bandgap. We present the variation of the optical bandgap of these samples with the Nb concentration ($x$) in Fig.~\ref{fig:DRS}(f). The bandgap value of the parent m-LaVO$_4$ ($x=$ 0) compound is about 3.4~eV, which is consistent with the reported experimental value of 3.5$\pm$0.2~eV in ref.~\cite{ParhiSSS08}. Moreover, the bandgap value decreases up to the $x=$ 0.1 sample to 3.22~eV and then almost constant for the samples with $x>$ 0.1 (3.24~eV). The optical absorption property of a semiconductor, which is relevant to the electronic structure serves as a key factor in determination of the photocatalytic activity \cite{HeJMCA11, MahapatraIECR07,MakulaJPCL18}.

\section*{\noindent ~Conclusions}

In conclusion, we have successfully prepared polycrystalline LaV$_{1-x}$Nb$_x$O$_4$ ($x=$ 0--0.2) samples by solid-state reaction method including the pure scheelite-tetragonal phase. The room temperature x-ray absorption measurements reveal that the La cation exists in a trivalent oxidation state, while V and Nb cations have their 5+ oxidation state in a tetrahedral coordination. The substitution of Nb ions in place of V ions reveals a structural phase transformation due to its larger size as compared to the V$^{5+}$. The Rietveld refinement of room temperature x-ray diffraction patterns demonstrate that the $x=$ 0 sample exist in a monoclinic (m) phase, whereas for the $x=$ 0.05--0.2, both monoclinic and scheelite-tetragonal (s-t) phases co-exist in a certain fraction. Interestingly, a monotonous enhancement in the Raman spectral intensity with Nb substitution is correlated with the substitution induced increase in the s-t phase. The room temperature photoluminescence measurements on these samples exhibit a broad spectra, the deconvolution of these spectra using six Gaussian peaks designate the availability of more than one electron-hole pairs recombination center. Moreover, the FTIR spectra indicate that the Nb substitution gives origin to some additional infrared active modes owing to the deformation of the VO$_4^{3-}$ tetrahedra. 

\section*{\noindent ~Acknowledgments}
H.D. and R.S. are grateful to MHRD and DST-INSPIRE, respectively for their financial support through fellowship. Authors thank FIST (DST, govt. of India) UFO scheme of IIT Delhi for providing Raman facility at physics department. We acknowledge the Central Research Facility (CRF) for EDX, Physics Department for XRD, and Nanoscale Research Facility (NRF) for PL, FTIR, and UV-Vis facilities. We thank Ravi Kumar and S. N. Jha for help and support during XAS measurements at RRCAT, India. R.S.D. acknowledges the financial support from SERB-DST through Early Career Research (ECR) Award (project no. ECR/2015/000159) and UGC-DAE CSR Indore through CRS project no. CSR-IC-ISUM-36/CRS-319/2019-20/1371. A high temperature furnace (from Nabertherm GmbH, Germany) was used for sample preparation, supported from BRNS through DAE Young Scientist Research Award (Project Sanction No. 34/20/12/2015/BRNS).

\end{document}